\documentclass[preprint]{aastex}
\usepackage{emulateapj5,psfig}
\usepackage{apjfonts}
\usepackage{natbib}

\newcommand{\kms}{km s$^{-1}$}
\newcommand{\msun}{$M_{\sun}$}
\newcommand{\lsun}{$L_{\sun}$}
\newcommand{\hubu}{km s$^{-1}$ Mpc$^{-1}$}

\newcommand{\ha}{H$\alpha$}
\newcommand{\hb}{H$\beta$}

\newcommand{\oiii}{{\rm [O~III]}}

\newcommand{\nii}{{\rm [N~II]}}
\newcommand{\sii}{{\rm [S~II]}}
\newcommand{\oi}{{\rm [O~I]}}

\def\wave#1{$\lambda${#1}}
\def\waves#1{$\lambda\lambda${#1}}

\slugcomment{Received 2002 October 11; accepted 2002 November 14}
\journalinfo{Accepted for publication in \it{The Astrophysical Journal}}

\shortauthors{Tran et al.}
\shorttitle{UGC 10214 Star Clusters}

\begin{document}

\title{Advanced Camera for Surveys Observations of Young Star Clusters in the Interacting Galaxy UGC 10214}

\author{H. D. Tran\altaffilmark{1}, M. Sirianni\altaffilmark{1},
H. C. Ford\altaffilmark{1}, G. D. Illingworth\altaffilmark{2},
M. Clampin\altaffilmark{3}, G. Hartig\altaffilmark{3},
R. H. Becker\altaffilmark{4}, R. L. White\altaffilmark{3},
F. Bartko\altaffilmark{5}, N. Ben\'{\i}tez\altaffilmark{1},
J. P. Blakeslee\altaffilmark{1}, R. Bouwens\altaffilmark{2},
T. J. Broadhurst\altaffilmark{6}, R. Brown\altaffilmark{3},
C. Burrows\altaffilmark{3}, E. Cheng\altaffilmark{7},
N. Cross\altaffilmark{1}, P. D. Feldman\altaffilmark{1},
M. Franx\altaffilmark{8}, D. A. Golimowski\altaffilmark{1},
C. Gronwall\altaffilmark{9}, L. Infante\altaffilmark{10},
R. A. Kimble\altaffilmark{7}, J. Krist\altaffilmark{3},
M. Lesser\altaffilmark{11}, D. Magee\altaffilmark{2},
A. R. Martel\altaffilmark{1}, Wm. J. McCann\altaffilmark{1},
G. R. Meurer\altaffilmark{1}, G. Miley\altaffilmark{8},
M. Postman\altaffilmark{3}, P. Rosati\altaffilmark{12},
W. B. Sparks\altaffilmark{3}, and Z. Tsvetanov\altaffilmark{1}}

\altaffiltext{1}{Department of Physics and Astronomy, Johns Hopkins University, Baltimore, MD 21218.}
\altaffiltext{2}{UCO/Lick Observatory, University of California, Santa Cruz, CA 95064.}
\altaffiltext{3}{STScI, 3700 San Martin Drive, Baltimore, MD 21218.}
\altaffiltext{4}{Physics Department, University of California, Davis, CA 95616, and Institute for Geophysics and Planetary Physics, 
L-413, Lawrence Livermore National Laboratory, 7000 East Avenue, Livermore, CA 94550.}
\altaffiltext{5}{Bartko Sci. \& Tech., P.O. Box 670, Mead, CO 80542-0670.}
\altaffiltext{6}{The Racah Institute of Physics, Hebrew University, Jerusalem 91904, ISRAEL.}
\altaffiltext{7}{NASA-GSFC, Greenbelt, MD 20771.}
\altaffiltext{8}{Leiden Observatory, P.O. Box 9513, 2300 Leiden, The Netherlands.}
\altaffiltext{9}{Department of Astronomy and Astrophysics, The Pennsylvania State University, 525 Davey Lab, University Park, PA 16802.}
\altaffiltext{10} {Pontificia Universidad Catolica de Chile, Santiago, Chile.}
\altaffiltext{11}{Steward Observatory, University of Arizona, Tucson, AZ 85721.}
\altaffiltext{12}{European Southern Observatory, Karl-Schwarzschild-Str. 2, D-85748 Garching, Germany.}


\begin{abstract}

We present the first {\it Advanced~Camera~for~Surveys} (ACS)
observations of young star clusters in the colliding/merging galaxy
UGC 10214. The observations were made as part of the Early Release
Observation (ERO) program for the newly installed ACS during service
mission SM3B for the {\it Hubble~Space~Telescope} ($HST$).  Many young
star clusters can be identified in the tails of UGC 10214, with ages
ranging from $\sim$ 3 Myr to 10 Myr.  The extreme blue $V-I$
(F606W$-$F814W) colors of the star clusters found in the tail of UGC
10214 can only be explained if strong emission lines are included with
a young stellar population. This has been confirmed by our Keck
spectroscopy of some of these bright blue stellar knots.  The most
luminous and largest of these blue knots has an absolute magnitude of
$M_V = -14.45$, with a half-light radius of 161 pc, and if it is a
single star cluster, would qualify as a super star cluster (SSC). 
Alternatively, it could be a superposition of multiple scaled OB associations
or clusters. 
With an estimated age of $\sim$ 4-5 Myr, its derived mass is $<$ 1.3
$\times 10^6$ \msun.  Thus the young stellar knot is unbound and will not
evolve into a normal globular cluster.  The bright blue clusters and
associations are much younger than the dynamical age of the tail,
providing strong evidence that star formation occurs in the tail long 
after it was ejected. UGC 10214 provides a nearby example
of processes that contributed to the formation of halos and
intra-cluster media in the distant and younger Universe.
\end{abstract}

\keywords{galaxies: individual (VV 29, Arp 188, UGC 10214) -- galaxies: star clusters}

\section{Introduction} \label{intro}

In 2002 April, as part of the early release observations (EROs) to
demonstrate the new capabilities of the Advanced Camera for Surveys
(ACS; Ford et al. 1998) installed in early March 2002 on the Hubble
Space Telescope ($HST$), deep images of the interacting galaxies NGC
4676 (``The Mice'') and UGC 10214 (``Tadpole'' = VV 29 = Arp 188) were
obtained, using the Wide Field Channel (WFC) with the F475W ($g$),
F606W (broad $V$) and F814W ($I$) filters.  NGC 4676 is a pair of
galaxies caught in the act of a collision, while UGC 10214 appears as
a disturbed spiral with a long tail of stars.  In this paper, we
present the observations, and focus on the search and identification
of very young star clusters found in the long tidal tail of one of
these galaxies, UGC 10214. In addition, we present optical
spectroscopy of some of these brightest clusters obtained with the
Keck 10m telescope. A detailed analysis of the data, including the
magnitudes, colors, specific frequencies and spatial distributions of
candidate clusters found in the both systems of interacting galaxies
will be the subject of a future paper (Sirianni et al. 2002).
Analysis of the numerous faint galaxies visible in the background
field will also be presented in a separate paper (Ben\'{\i}tez et
al. 2002). The redshift of UGC 10214 is $z = 0.03136$ \citep{brig01},
placing it at a distance of 125 Mpc ($H_o = 75$ \hubu). At this
distance, 1\arcsec~corresponds to 606 pc, and 1 WFC pixel covers 30.3 pc.

\section{Observations and Reduction} \label{obs}
\subsection{ACS Imaging} \label{obsacs}

The initial ACS observations of UGC 10214, dubbed the Tadpole, were
obtained on 2002 April 1 (UT), using the ACS/WFC through the F475W
($g$), F606W (broad $V$) and F814W ($I$) filters.  We used POS TARG
command offsets of 3\arcsec~in the Y direction to bridge the gap
between the two WFC chips.  In order to minimize the visibility of the
gap, for each filter, we commanded the telescope offsets to three
pointings, at (0\arcsec, 0\arcsec), (0\farcs248, 3\farcs001) and
($-$0\farcs248, $-$3\farcs001).  At each POS TARG position, two
individual exposures were taken to aid in the removal of cosmic rays.
Unfortunately, there was an error in the telescope pointing for the
first set of observations, which caused part of the ``head'' of the
Tadpole to get cut off from the field of view.  A second set of
observations were subsequently obtained on 2002 April 9, using the
identical settings and sequence of exposures as the first, but with a
slight change in the position and telescope orientation to include the
entire head and body of the ``Tadpole'' in the full WFC field of view.
The total exposure times for the sum of both sets of observations were
13,600s (6 orbits) in F475W, 8,040s (4 orbits) in F606W, and 8,180s (4
orbits) in F814W.

The data were reduced by first calibrating with CALACS, which included
bias and dark subtraction, flat fielding, and cosmic ray rejection,
then processed through the ACS Investigation Definition Team pipeline 
at Johns Hopkins University, which measured offsets and rotations between
individual dithered images from both datasets and combined them 
while removing geometric distortion, residual cosmic rays and
detector defects. We used the best reference files available at
the time of reduction.  The reader is referred to the ACS Data Handbook 
(Mack et al. 2002) for details.

Identification of point-like objects (stars and clusters) and their
photometry were performed using DAOPHOT with an aperture radius of 3
pixels (0\farcs15) and sky background between 10-15 pixels. Since the
size of a WFC pixel (30.3 pc) is much larger than the typical
effective radius of young blue clusters found in nearby galaxies
\citep{lar99}, the majority of the compact clusters identified in our
image are not resolved.  The aperture correction was therefore
calculated by measuring the encircled energy profiles for three
isolated stars in the field. The corrections applied were: 0.11
(F475W), 0.14 (F606W) and 0.17 (F814W). All magnitudes are in AB
system with the following zero points: 25.87 (F475W), 26.36 (F606W)
and 25.86 (F814W). No correction for Galactic extinction was
performed, as it is negligible. We used the following selection
criteria to identify the cluster candidates: i) cluster must be
present in all three bands, ii) FWHM $>$ 1 pixel in all 3 bands, iii)
$M_V < -9.0$, to exclude stars, and iv) magnitude error $< 0.3$ in all
three bands.

\subsection{Keck Spectroscopy} \label{obskeck}

Long-slit spectra were obtained on 2002 May 11 with the Echellette
Spectrograph and Imager (ESI, Sheinis et al. 2002) on the Keck II
telescope using the multiorder echellette mode with an 1\arcsec~wide
slit.  The slit was centered on a cluster or ``knot'' of bright blue
stars in the tail of UGC 10214 [RA=16 06 15.16, Dec= +55 25 48.74
(2000)], and oriented at PA = 70\arcdeg~along the elongation of the
blue knot to include a trail of nearby clumps of stars (see
Fig. \ref{bluecl}).  A single 900s exposure was obtained through
variable clouds, giving a wavelength coverage from 3900 \AA~to 11000
\AA, with a two pixel spectral resolution of $\sim$ 55 \kms
(FWHM). The data were reduced using standard IRAF tasks.

\section{Results} \label{res}

\begin{figure*}[t]
\centerline{\psfig{file=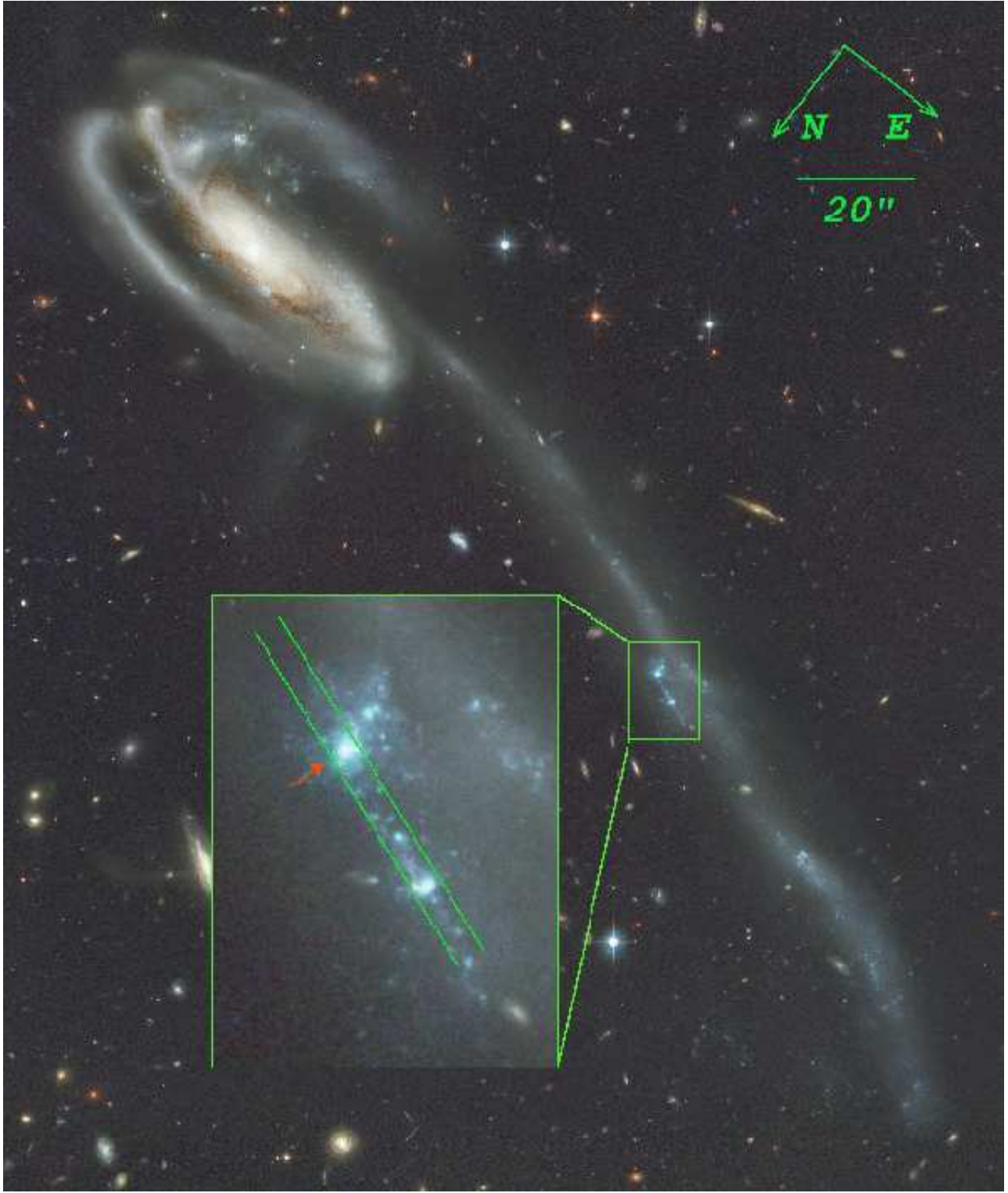,width=17.0cm,angle=0}}
\caption{Composite $g$, $V$, and $I$ color image of UGC 10214, with a
close-up of the bright blue clump in the tail. The brightest stellar
knot marked by the red arrow is a probable super star cluster
(SSC). Parallel lines indicate the slit width and orientation used in
the Keck spectroscopy. \label{bluecl}}
\end{figure*}

Figure \ref{bluecl} shows a color image of UGC 10214 and a close-up of
the bright extended stellar association found in the tail of the galaxy. 
We identified a total of 42 cluster candidates in this region of the tail; 
their F475W $-$ F606W ($g-V$) vs. F606W $-$ F814W ($V-I$) color-color 
diagram is shown in Figure \ref{clumpcc}. The points generally cluster
in a narrow range of $g-V$ color, but 

\psfig{file=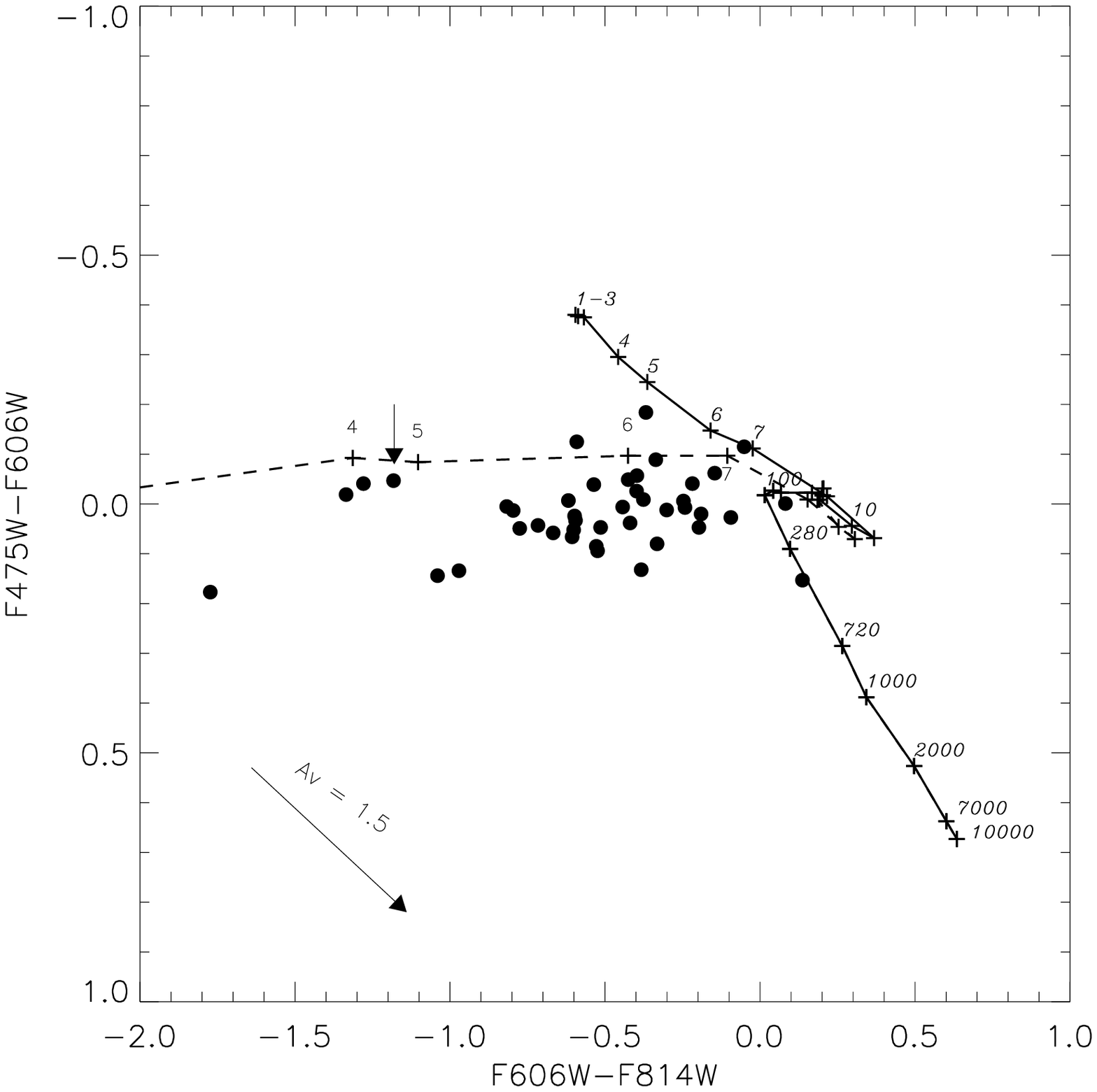,width=8.9cm}
\figcaption{F475W $-$ F606W ($g-V$) vs. F606W $-$ F814W ($V-I$)
color-color diagram for the 42 clusters identified in the bright blue
extended clump in the tail of UGC 10214. The solid curve shows the
stellar evolutionary track from the population synthesis models (for
instantaneous starburst with Salpeter IMF, and empirical stellar
spectra with solar metallicity) of Bruzual \& Charlot (2001).  The
dashed curve denotes the same model but with the addition of emission
lines. Labels along the tracks indicate ages in Myr. A reddening
vector for $A_V = 1.5$ is also shown.  The arrow marks the brightest
stellar knot (red arrow in Fig. \ref{bluecl}), which has an
absolute magnitude of $M_V = -14.45$ and qualifies as a super star
cluster (SSC). The extreme blue $V-I$ color of this and other clusters
cannot be explained by the stellar population alone, but is due
instead to the presence of strong emission lines. \label{clumpcc}}
\vskip 0.3cm

\noindent
they cover a wider range in $V-I$, from zero to $\approx$ 2.0.  

The brightest stellar knot (marked by an arrow in Fig. \ref{bluecl})
of the extended blue clump is very blue ($V-I$ = $-$1.18, $g-V$ =
$-$0.05), and has apparent AB magnitudes of F475W ($g$) = 21.07, F606W
($V$) = 21.12, F814W ($I$) = 22.30. With an absolute magnitude of $M_V
= -14.45$, it is most likely a very young massive super star cluster
(SSC).  The combination of the moderate $g-V$ color but extremely blue
$V-I$ color cannot be explained by the usual population synthesis
models of Bruzual \& Charlot (2001, hereafter BC01) or Starburst99
models of Leitherer et al. (1999, hereafter SB99), even for stars as
young as 1 Myr (see Fig. \ref{clumpcc}). The simplest way that these
colors could be fitted is to include strong emission lines in the
starburst models of BC01 and SB99. The alteration of colors of young
compact star clusters due to the presence of strong emission lines has
been recognized for some time, as discussed in e.g., \citet{mk91},
\citet{may95}, \citet{sti98} and \citet{wz02}, among others.

Shown in Figure \ref{modsp} are the model spectrum of such a cluster.
The model was constructed from a combination of the stellar continuum
of BC01 and the strongest and most common emission lines, whose
strengths were derived based on the \hb~equivalent width versus age
relationship of SB99, and assuming typical line ratios for a starburst
region.  Also plotted in the figure are the filter bandpasses for the
three filters used in the observation. As can be seen, for the
redshift of UGC 10214, the F606W filter would allow all of the
strongest emission lines through, while they are all excluded from the
F814W filter. On the other hand, the strong emission lines of
\oiii~\waves 4959, 5007, but not the \ha~+ [N II] complex, are allowed
through the F475W filter. This results in a $V-I$ (F606W$-$F814W)
color that is too blue compared to a stellar continuum from a
pure (i.e., no emission line) population synthesis model, but a $g-V$
(F475W$-$F606W) color that is not unusual.  This clearly demonstrates
that the addition of emission lines could naturally explain the
extreme $V-I$ observed color, while the $g-V$ color remained relatively unchanged.

The color-color evolutionary tracks derived from the model described
above is shown in Figure \ref{clumpcc}. As can be seen, the model with
emission lines (dashed curve) appears to fit better with the observed
data points than that without (solid curve).  Most of the candidate
clusters lie close to and slightly below the 

\psfig{file=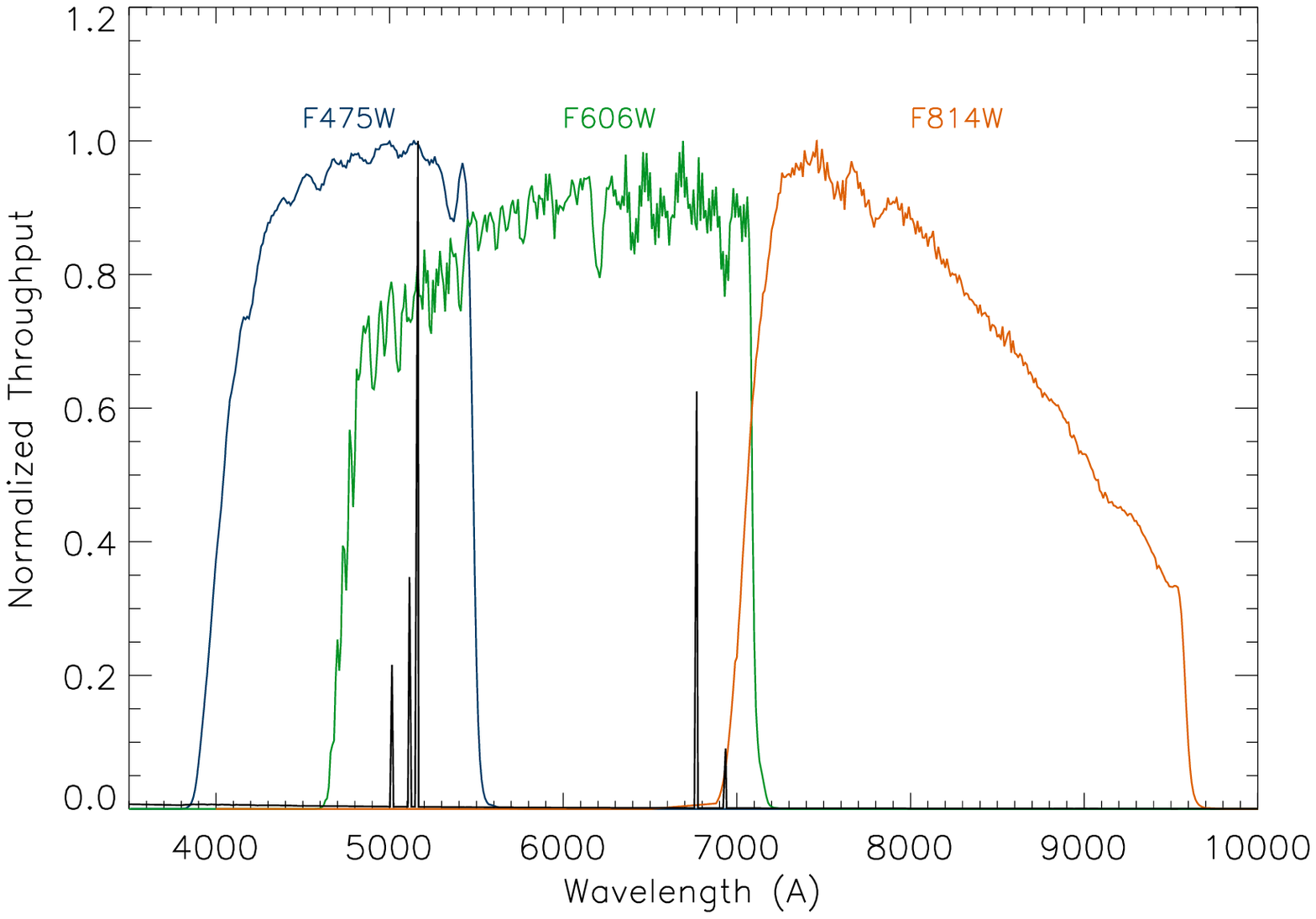,width=9.0cm,height=7.0cm}
\figcaption{Model optical spectrum of the SSC in the tail of UGC 10214. Overlaid
are the transmission curves of the three filters used. At the redshift of 
UGC 10214, most strong emission lines are able to pass through both F475W 
and F606W filters, while none is allowed through F814W. \label{modsp}}
\vskip 0.3cm

\begin{figure*}[t]
\centerline{\psfig{file=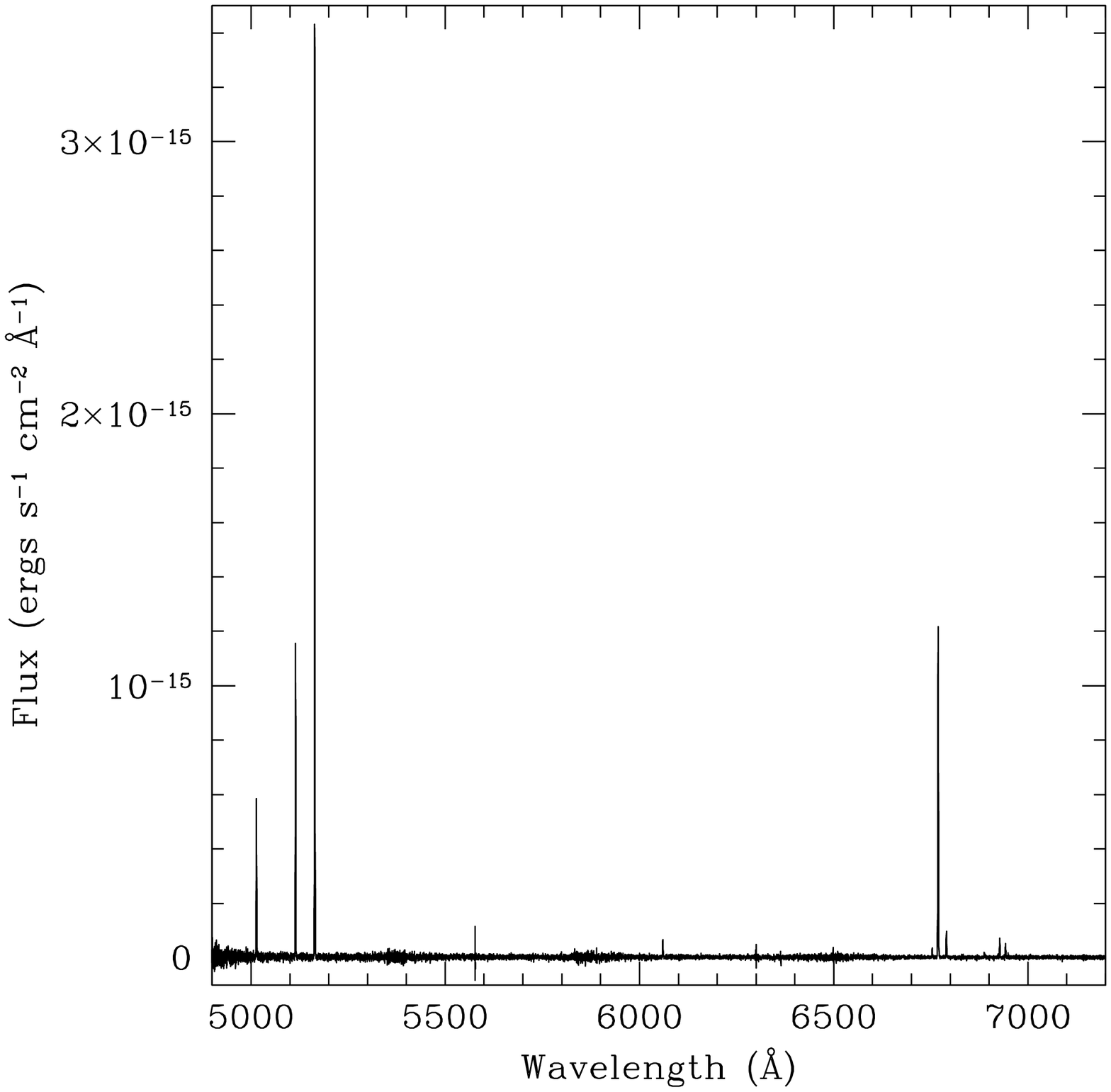,width=13.0cm,angle=0}}
\caption{Keck observed ESI spectrum of the brightest blue stellar knot
(SSC) in the tail of UGC 10214. Strong \hb~and \oiii~\waves 4959, 5007
emission lines are seen to the left around 5100 \AA, and
\ha+\nii~complex is at near 6750 \AA. \label{sscsp}}
\end{figure*}

\noindent
modeled track, and are consistent
with ages in the range 3-10 Myr.  The slightly redder $g-V$ color of
most of the older clusters compared to the modeled curve could be explained
by a small amount of extinction.  From a comparison with the model, 
the best age estimate for the SSC is about 4-5 Myr.

Based on these findings, in May 2002 we obtained Keck spectroscopy of
the blue knot (see \S \ref{obskeck}) in order to test the following:
i) confirm the presence of strong emission lines as required by the
models.  ii) look for the expected signatures of Wolf-Rayet (W-R) stars
which descended from the very massive O stars ($\ga$ 30 \msun) in the
cluster, and iii) better constrain the spectral energy distribution
(SED) of the star cluster and test the theoretical models of BC01 and SB99.

A Keck spectrum of the brightest blue knot is shown in Figure \ref{sscsp}.  
As predicted, very strong emission lines are seen. These lines include
those of \hb, \ha, \oiii~\waves 4959, 5007, \nii~\waves 6548, 6583,
and \sii~\waves 6717, 6731. The heliocentric velocity measured from
these lines is 9422 $\pm$ 7 \kms. This is consistent with the HI velocity
measured by \citet{brig01} at the position of this knot. 
The emission line widths are very narrow, but
may be slightly resolved in our spectra. The observed FWHM is $\approx$
75 \kms, giving an instrumental-corrected (subtraction in quadrature) 
width of $\sim$ 50 \kms. The observed Balmer decrement \ha/\hb~is 2.76,
indicating very little or no reddening toward the tail of UGC 10214 at
the position of this blue knot.  Other line ratios are: \oiii~\wave
5007/\hb~= 5.53, \nii~\wave 6583/\ha~=0.0705, \sii~(6717 + 6731)/\ha~=
0.090, \oi~\wave 6300/\ha~= 0.012, \oiii~(5007 + 4959)/4363 = 86, and
\sii~6717/6731 = 1.35.  These emission-line ratios yield an electron
temperature of about 1.36 $\times 10^4$ K and density of 60 cm$^{-3}$,
fairly typical of photoionized gas in metal-poor H II regions. They also 
place them among areas typically occupied by H II regions and H II galaxies
in the diagnostic diagrams of \citet{vo87}. 

The observed optical continuum is very faint and barely
detectable. This results in a very uncertain determination of the
equivalent width (EW) of the \hb~emission line.  We measure an \hb~EW of
290 $\pm$ 255 \AA, the error for which is dominated by that of the continuum
level. In principle, this can be used to constrain
the age of the star cluster (SB99; Schaerer \& Vacca 1998), yielding
an estimate of 1-6 Myr, consistent with the age obtained from
population synthesis model fits to ACS photometry. Due to the
faintness of the continuum and the non-optimum observing conditions,
the S/N is insufficient to allow a search for W-R features, such as
the 4686\AA~bump, which is expected for a young star cluster of this
age. Better quality spectra would be needed to confirm the existence
of W-R stars and constrain the star formation history of the cluster.

\section{Discussion} \label{disc}

It is of interest to ask if the large knots or clumps of blue stars
found in the tail of UGC 10214 were created from the galactic
encounter and whether they are precursors to the compact globular
clusters (GC) that are found in our Galaxy and essentially all
galactic halos. These star clusters are similar to those found in the
tidal debris of the Antennae Galaxies (NGC 4038/4039, e.g., Whitmore et
al. 1999) or of the Stephan's Quintet (NGC 7318/7319, Gallagher et al. 2001).
 
We can estimate the mass of the brightest blue stellar ``knot'', or
SSC in the tail of UGC 10214 from its estimated age and observed
brightness by using an existing relationship between age and
mass-to-light ratio ($M/L$) for young stellar clusters derived by
\citet{chan99} using Padua models (Chiosi, Vallenari, \& Bressan
1997). With a derived age of $\sim$ 4-5 Myr, the correlation gives log
$M/L_V = -1.6$. The observed absolute magnitude for the SSC in F606W
is $-$14.45 or $L_V = 10^{7.7}$ \lsun, yielding an estimated mass of
1.3 $\times 10^6$ \msun. We note that this actually represents an
upper limit of the mass since the modeled $M/L$ does not include
contribution from emission lines. 
To correct for the emission-line contribution, we compared the integrated 
flux of our modeled spectrum with and without emission lines through the 
F606W filter. The difference in cluster brightness
is found to be 0.75 mag in F606W, giving a corrected (line-free) 
absolute magnitude of $-$13.70 or $L_V = 10^{7.42}$ \lsun, and reducing 
the mass to 6.6 $\times 10^5$ \msun. These values are typical for a SSC. 
In particular, the luminosity and mass of this SSC are similar to
those of the $\sim$ 60 Myr old SSC ``F'' in M82 \citep{sg01}, or of the 
15 Myr old SSC ``1447'' in NGC 6946 \citep{lar01}.

The 3 to 10 Myr ages of the clusters in the tidal tail show that they
formed close to their present positions.  The projected distance of
the $\sim$ 4.5 Myr old SSC is 61 kpc from the center of UGC 10214.  
Assuming that the maximum ejection velocity available to the cluster is 
comparable to the 400 \kms~rotation speed measured by \citet{brig01}
for the main galaxy, the dynamical age for the tail is at least 150 Myr old. 
Since this is much older than the estimated age of the SSC, we conclude
that the young clusters in the tail formed long after the galactic encounter
that produced the tidal debris. 
This conclusion is consistent with the mechanism proposed
by \citet{w90}, who suggested that star formation was triggered 
in the tail by density enhancements produced from tidal forces in
the collision.  
It is also consistent with the suggestion by \citet{HvG96} that ``gaps''
(region of reduced surface brightness) in tidal tails are induced by 
self-gravity in the tails. The two most prominent regions of star formation 
in UGC 10214 tail are separated by the most conspicuous gap in the tail. 
In situ star formation in tidal tails has also been observed 
in many other systems (e.g., Schweizer 1978; Mirabel, Dottori, \& Lutz 1992),
and appears to be a common feature of such tails.

Fitting the surface brightness profiles of the SSC with a king model
\citep{king62} gives a core radius of about 0\farcs155, or 94 pc,
corresponding to a half light radius of $r_e$ = 0\farcs266~= 161
pc. To our knowledge, the largest $r_e$ measured for a SSC or a scaled
OB association (SOBA) is 84 pc \citep{mai01}.  As a comparison, the
F-M82 SSC, which has similar mass and luminosity, has a half light
radius of 2.8 pc.  Therefore, the unusually large size of the SSC in
UGC 10214 suggests that this cluster could be a particularly populous
OB association or an unresolved superposition of a multiple system
like that in NGC 1569 A1-A2 \citep{dmar97}.  In contrast to the case
of 30 Doradus Nebula, which has a very large emission-line region but 
the stellar component is largely confined to the very compact R136 star 
cluster \citep{wmb02}, we can rule out the possibility of a similar 
significant nebula contribution to the large size of the UGC 10214 SSC.  
The UGC 10214 cluster radius measured from the F814W data, where stellar 
continuum dominates, is not significantly different from those measured 
from the F475W and F606W data, where line emission dominates. Thus, the 
measured size of the SSC in UGC 10214 can be taken as genuinely due mostly 
to stars.

The above estimated mass of the SSC (6.6 $\times 10^5$ \msun) is
$\approx$~3 $\times$ the mass of a typical globular cluster (GC;
Mandushev, Spassova, \& Staneva 1991), while its radius is
approximately 50 $\times$ the size of a GC. This corresponds to a mass 
density that is at least 4 orders of magnitude smaller than that of a GC. 
Assuming that the projected half-mass radius $r_m =  4/3~r_e$ \citep{spi87},
the mass density within the half-mass radius is only $\sim 10^{-2}$
$M_\odot$ $pc^{-3}$.
Thus, under the assumption that the SSC represents
a single cluster of young stars, it appears to be quite ``fluffy'' for
its mass.  For an isotropic, gravitationally bound system, the total
cluster mass is given by the virial theorem: $ M = 3\sigma^2 R/G$,
where $\sigma$ is the line-of-sight velocity dispersion and $R$ is the
effective radius. Therefore, to remain bound like a GC, $\sigma$ for
the blue SSC has to be $\lesssim 1/4 \times \sigma_{GC}$. The
velocity dispersion of a Galactic GC is typically less than 
20 \kms, with a mean of $\sigma_{GC} = 8.4$ \kms~(Mandushev et al. 1991), 
which means that the stellar velocities in the SSC cannot be greater than
$\sim$ 5 \kms, if it is to remain bound. This cannot be the case, if the 
stars move about like the gas ($\sim$ 50 \kms, \S \ref{res}) in the cluster.
Thus the blue clump in the tail of UGC 10214 is unlikely grow up to be a GC, 
unless it is dynamically very cold.

Because we do not know the distribution of mass in UGC 10214's halo
and the geometry of the tidal tail, it is difficult to say whether or
not the stars and gas in the tail are gravitationally bound to the
parent galaxy. \citet{brig01} found that the HI velocities in the tail
are only $\sim$100 \kms~higher than the HI systemic velocity of UGC 10214,
with little or no velocity gradient over the 100 kpc length of the
tail. They also revealed a possible counter tail on the western side
of the main galaxy with an approaching velocity of $\sim$ 300 \kms.
Such tail kinematics are like those of other mergers sporting
similarly long tidal tails (e.g., NGC 7252, Hibbard \& Mihos 1995; Arp
299, Hibbard \& Yun 1999).  Numerical simulations of NGC 7252
\citep{hm95} have indicated that most of the tail material will remain
bound, falling back towards the merger remnant over several billion
years, perhaps forming rings, shells and high-velocity streams.
Similar detailed modeling for UGC 10214 should be very valuable in
accounting for its tidal morphology and tail kinematics, and predicting
their future evolution, but it seems likely that the stars and gas
will remain bound in the poor cluster WBL 608 \citep{W99} to which UGC
10214 belongs.  UGC 10214 and galaxies like it provide nearby examples
of processes that contributed to the formation of halos and
intra-cluster media in the distant and younger Universe, when the
density of galaxies was higher.
   
\section{Conclusions} \label{conc}

We obtained ACS WFC observations of the peculiar merging galaxy UGC
10214, and presented Keck optical spectra of a luminous, young blue
stellar clump located in the tail of the galaxy. The unusually blue
$V-I$ color but normal $g-V$ color of the clump can be well explained
by the presence of strong emission lines characteristic of H II
regions embedded in a super cluster of massive, young
stars. Population synthesis models indicate the age of the star
cluster to be $\sim$~4-5 Myr. The large size of the cluster, combined
with its moderate estimated mass of $M \sim 6.6 \times 10^5$
\msun~suggest that its mass density is too low to allow it to
evolve into a normal globular cluster.
The bright blue clusters and associations lie at large
projected distances, with an age much younger than that of the tail,
providing strong evidence that star formation occurs
in the tail long after it was ejected.

\acknowledgments 
ACS was developed under NASA contract NAS5-32864, and
this research is supported by NASA grant NAG5-7697. We are grateful
for an equipment grant from the Sun Microsystems Corporation.  The
W. M. Keck Observatory is operated as a scientific partnership between
the California Institute of Technology and the University of
California, made possible by the generous financial support of the
W. M. Keck Foundation. Work performed at the Lawrence Livermore
National Laboratory is supported by the DOE under contract
W7405-ENG-48. We thank an anonymous referee for several useful comments.


\end{document}